%% file: main.tex
\documentclass[fullpage]{article}
\usepackage{graphicx} 
\usepackage{xcolor}
\usepackage{url}
\usepackage[textsize=scriptsize]{todonotes}

\begin{document}

{\centerline {\bf {\Large Science of Safe AI}}}

\vspace*{.1in}

\begin{center}
    {\bf Report based on NSF Workshop}\\
    {\bf at University of Pennsylvania, Philadelphia, on February 26, 2025}
\end{center}

\vspace*{.1in}

\noindent
{\bf Workshop Organizer:}
Rajeev Alur (University of Pennsylvania).

\vspace*{.1in}
\noindent
{\bf Discussion Leads:}
Corina P\u{a}s\u{a}reanu (NASA Ames and CMU), 
Greg Durrett (The University of Texas at Austin),
Hadas Kress-Gazit (Cornell University), and
Ren\'e Vidal (University of Pennsylvania).

\vspace*{.1in}
\noindent
{\bf Participants:}
Aaron Roth (Penn),
Aditya Akella (UT Austin),
Andrea Bajcsy (CMU),
Andreas Malikopoulos (Cornell),
Anindya Banerjee (NSF),
Anushri Dixit (UCLA),
Armando Solar-Lezama (MIT),
Avi Shinnar (IBM Research),
Cang Ye (NSF),
Chaowei Xiao (Wisconsin, Madison),
Cho-Jui Hsieh (UCLA),
Christopher Yang (NSF),
Claire Tomlin (UC Berkeley),
Dan Roth (Penn),
Daniel Brown (Utah),
Dinesh Jayaraman (Penn),
Dung Tran (Univ. Nebraska),
Eric Atkinson (Binghamton),
Esin Tureci (Princeton),
Feras Saad (CMU),
George Pappas (Penn),
Hamed Hassani (Penn),
Han Zhao (UIUC),
Hanghang Tong (UIUC),
Hanlin Zhang (Harvard),
Haoze Wu (Amherst College),
Huan Zhang (UIUC),
Insup Lee (Penn),
Jaime Fernandez Fisac (Princeton),
Jan Hoffmann (CMU),
Jia Deng (Princeton),
Jie Yang (NSF),
Joydeep Biswas (UT Austin),
Kai Shu (Emory),
Kai-Wei Chang (UCLA),
Kyriakos Vamvoudakis (Georgia Tech),
Lars Lindemann (USC),
Loris D'Antoni (UC San Diego),
Madhur Behl (Virginia),
Madhusudan Parthasarathy (UIUC),
Mahdi Khalili (Ohio State),
Mayur Naik (Penn),
Michael Littman (NSF),
Ming Jin (Virginia Tech),
Mohammad Fadiheh (Stanford),
Mohit Bansal (UNC Chapel Hill),
Momotaz Begum (New Hampshire),
Moshe Vardi (Rice),
Naira Hovakimyan (UIUC),
Nikolai Matni (Penn),
Olga Russakovsky (Princeton),
Osbert Bastani (Penn),
Ruzena Bajcsy (Penn),
Sandhya Saisubramanian (Oregon State),
Sanghamitra Dutta (Maryland),
Sayan Mitra (UIUC),
Shangtong Zhang (Virginia),
Sharon Li (Wisconsin, Madison),
Shenlong Wang (UIUC),
Shreyas Kousik (Georgia Tech),
Sorin Draghici (NSF),
Sourya Dey (Galois),
Steven Holtzen (Northeastern),
Surbhi Goel (Penn),
Tang Li (Delaware),
Taylor Johnson (Vanderbilt),
ThanhVu Nguyen (George Mason),
Thema Monroe-White (George Mason),
Tianyi Zhang (Purdue),
Weiming Xiang (Augusta),
Xi Peng (Univ. Delaware),
Xian Yu (Ohio State),
Xiaofeng Wang (South Carolina),
Xueru Zhang (Ohio State),
Yongming Liu (Arizona State),
Ziwei Zhu (George Mason),
and Ziyu Yao (George Mason).

\newpage
\section*{Executive Summary}

\input{summary}

\newpage
\section{Introduction}
\input{intro}

\section{Defining Safety}
\input{safety}

\section{Design for Safety}
\input{design}

\section{Safety Analysis}
\input{analysis}

\section{Attacks and Defenses}
\input{llm}

\newpage
\bibliographystyle{abbrv}
\bibliography{refs}
\end{document}

%% file: summary.tex
Recent advances in machine learning, particularly the emergence of foundation models, are leading to new opportunities to develop technology-based solutions to societal problems.
However, the reasoning and inner workings of today's complex AI models are not transparent to the user, and there are no safety guarantees regarding their predictions. 
Consequently, to fulill the promise of AI, we must address the following scientific challenge: how to develop AI-based systems that are not only accurate and performant but also safe and trustworthy?

The criticality of safe operation is particularly evident for autonomous systems for control and robotics, and was the catalyst for the Safe Learning Enabled Systems (SLES) program at NSF.
For the broader class of AI applications, such as users interacting with chatbots and clinicians receiving treatment recommendations, safety is, while no less important, less well-defined with context-dependent interpretations.
This motivated the organization of a day-long workshop, held at University of Pennsylvania on February 26, 2025, to bring together investigators funded by the NSF SLES program with a broader pool of researchers studying AI safety.
This report is the result of the discussions in the working groups that addressed different aspects of safety at the workshop.
The report articulates a new research agenda focused on developing theory, methods, and tools that will provide the foundations of the next generation of AI-enabled systems. 

\paragraph{Safety-First Agenda:}
Traditionally, the key design requirement for learning algorithms has been high accuracy.
The science of safe AI should make safety a first-class design objective in learning algorithms and architectures.
Key questions for system design are then: What is the suitable definition of safety in a given application context?
How do we integrate these safety requirements in design of learning algorithms?
How do analysis tools check whether a system satisfies such safety requirements?
What are the possible attacks on systems to make them unsafe and how do we defend against such attacks?
These are yet unsolved and challenging problems requiring long-term research.
Given the difficulty to monetize the value of safety, it is less of a priority for the AI industry.
This makes AI safety a particularly worthwhile focus for academic research funded by government agencies.

\paragraph{Cross-Disciplinary Methods:}
Researchers from 
 different communities are just beginning to address this challenge from their own perspectives.
Arguably, effective principles and tools for AI safety will bring together ideas from four distinct communities:
(1) {\bf human-AI interaction } researchers addressing what safety guarantees are desired for adequate levels of trust in
different application contexts, 
(2) {\bf ML theory} researchers designing new learning algorithms and theoretical tools for safety assurance,
(3) {\bf foundation models} researchers designing, training, and deploying new generations of learning systems with integrated safety objectives and defenses against potential attacks, and  
(4) {\bf formal methods} researchers developing analysis tools for
verifying that systems meet these requirements.

\begin{figure*}[h]
\begin{center}
    \includegraphics[width=5in]{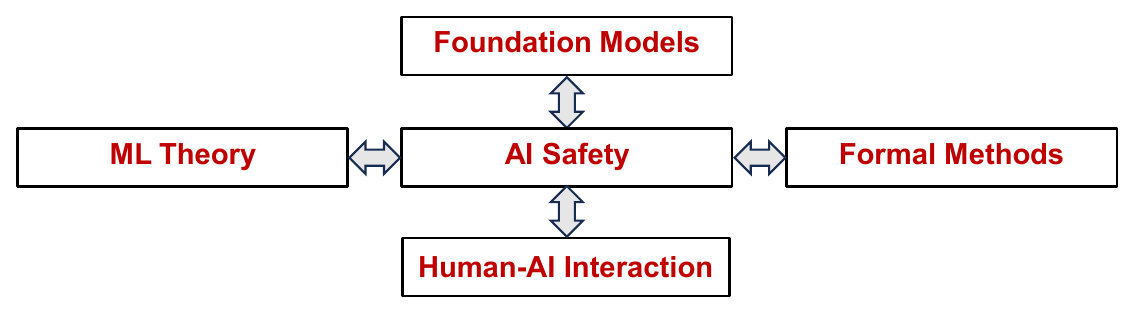}
\end{center}
\end{figure*}

A research program focused on AI safety will be a catalyst to foster the cross-disciplinary and cross-community collaboration necessary to address the safety challenge.

\paragraph{Research Directions:}
There is a rich variety of promising technical directions that can potentially contribute to AI safety research.
Representative problems include: 
How to define safety requirements that are both context dependent and computationally analyzable? 
How to rigorously quantify the uncertainty associated with predictions by learning algorithms?
How to integrate symbolic reasoning in neural architectures to improve assurance guarantees? 
How to develop monitoring algorithms to audit AI-based systems for adherence  to formal requirements? 
How should safety be analyzed and communicated with different people with varying levels of understanding of the system? 
What kinds of attacks and defenses need to be understood for ``AI agents'' and long-horizon reasoning models like DeepSeek-R1? 
What role does model interpretability play in helping defend against attacks on LLM systems?
How to design comprehensive benchmarks that can be used to evaluate safety of AI systems in different contexts?
How to certify AI systems for different types of safety guarantees?

%% file: intro.tex
Recent advances in machine learning are leading to novel AI-based solutions to challenging computational problems. Yet, the state-of-the-art models do not provide adequate safety guarantees, and can make occasional mistakes on even simple problems. 
Indeed, for every headline capturing the public imagination about the promise of AI,
there is a cautionary headline about its vulnerabilities.
The resulting lack of trust is a daunting obstacle in adoption of AI-based systems in high-stakes settings such as autonomous robots and clinical decision making.
Developing principles, methods, and tools to ensure safety of AI-based systems, thus, is critical to realize the promise of AI for technology-based solutions to problems of societal importance.

Ensuring safety of autonomous systems requires more than improving accuracy, efficiency, and scalability: it requires ensuring that systems are robust to unexpected situations, and monitoring them for anomalous or unsafe behavior.
This motivated the launch of the Safe Learning-Enabled Systems (SLES) program by US National Science Foundation, in partnership with Open Philanthropy and Good Ventures. 
In the first two iterations of the call for proposals in 2023 and 2024, the program 
selected investigators for research to advance rigorous approaches to safety of autonomous systems that incorporate learning algorithms.
This report is the result of a workshop organized to bring together 
these investigators to discuss results of their research. 
The workshop was held at University of Pennsylvania in Philadelphia on February 26, 2025, hosted by 
ASSET --- Penn Engineering's Center for Trustworthy AI.

 Most of the SLES investigators have their roots in control theory, cyber-physical systems, or formal methods. An important goal of the workshop was to bring additional perspectives to AI safety.
 We identified the following research areas also as critical to the safety of AI systems: (1) theory aimed at improving design-time guarantees of learning algorithms and (2) enhancements to foundation models aimed at improving factuality and logical reasoning. Consequently, many researchers in  machine learning theory and natural language processing  focused on these problems were also invited to participate in the meeting.

The workshop program included invited talks and poster presentations by SLES investigators. Invited speakers were chosen to represent the breadth of topics relevant to AI safety: 
Aaron Roth (Penn) on quantifying uncertainty for predictions by AI models, Olga Russakovsky (Princeton) on interplay among data, models, and society, and Moshe Vardi (Rice) on safety verification of systems that include learning components.

The core of the workshop was centered around discussion by participants organized in the following four working groups:
\begin{enumerate}
    \item {\bf Defining Safety.} The discussion focused on the following questions: What type of safety assurance do people need when systems are deployed in everyday use vs. in critical decision-making? What are examples of safety properties that can be formalized rigorously? What are opportunities and challenges in context of existing research?
    \item {\bf Design for Safety.}   The discussion focused on the following questions: What are the current trends in incorporating safety as a design goal, in addition to accuracy, during training? Which aspects are currently unexplored, and what are the new opportunities?
    \item {\bf Safety Analysis.} This group considered the following questions: What type of formal guarantees are possible using verification, testing, and monitoring tools? What are the trade-offs between worst-case guarantees and statistical guarantees? What are the new challenges and opportunities? 
    \item {\bf Attacks and Defenses.} Discussion questions were: What are new types of attacks on LLMs, VLMs, and AI agents? What are the potential defenses?
\end{enumerate}

The subsequent sections of this report summarize the discussion in each of these four working groups, identifying limitations of the current techniques and opportunities for future research.

%% file: safety.tex
Safety is a term used often in the context of AI and autonomous systems; however, 
there does not exist one concise
definition for what people mean by ``safety". Here, we first describe one way to taxonomize the different notions of safety, then discuss how context affects the definition of safety, and then outline challenges and opportunities for future work.

\subsection{Types of safety}
\label{sec:types_of_safety}

We can roughly group the different definitions of safety that were discussed by the participants into three  categories; safety that is with respect to the system state, with respect to adversaries, and with respect to people interacting with the system.

\paragraph{Safety with respect to system states:} Safety here can be defined at different levels of abstraction, from grounded explicit constraints such as ``an autonomous car never drives above the speed limit", to abstract notions such as ``bad things do not happen". 

On the grounded side, the community has considered safety as: (i) definitions of safe sets in the state space of the system, for example a quadrotor does not hit the ceiling, (ii) reach-avoid sets and different temporal logics that include constraints, goals and reactions to events, (iii) absence of outliers, and (iv) empirical measures, such as miles driven without accident. 

On the more abstract side, participants defined safety as ``avoiding harm" and ``avoiding catastrophic failures". There was discussion regarding avoiding small inconveniences that compound into catastrophic failure. 

\paragraph{Safety with respect to adversaries:} Here, the discussion identified robustness and resilience as key elements of safety. A system is considered safe if it is resilient to adversarial perturbations, for example in the context of neural network verification. Another view is defining safety with respect to an attack model. 

\paragraph{Safety with respect to people:} Safety of AI systems, be they virtual or physical, has to be addressed from the perspective of the individual people and communities  interacting with them. A system can be safe based on the above definitions, but still be perceived as unsafe by people. Here, safety is tied to explainability and transparency; a system is safe if it can reason about and  communicate its expected behavior, its own reasoning, and its limitations. For example, ML models that give a ``correct" prediction based on ``wrong" reasons should be considered unsafe.  

\subsection{Safety depends on context}
Despite the different definitions of safety that emerged during the discussion, there was consensus among participants that safety and the degree of harm that may result from unsafe actions are context dependent; the same action may be a minor inconvenience in one context, and a catastrophic failure in another. A few representative aspects include:
\begin{itemize}
    \item Who the system interacts with: An autonomous robot spilling water in a lab is less harmful than the same robot spilling water in the house of a person needing assistance.
    \item Scale: individual vs. societal harm; one bad driver will cause less harm than a fleet of autonomous vehicles that all choose unsafe actions due to a problem with their decision-making AI.
    \item Type of system: safety is a concept that is used for hardware, software, and overall systems; in programming languages it could mean type safety, in generative AI it could mean no hallucinations, and in autonomous aircrafts it could mean collision avoidance.    
\end{itemize}

\subsection{Safety standards}
Participants brought up that there are different efforts going on around defining standards for safety in different contexts. These include:
\begin{enumerate}
    \item Autonomous driving: UL 4600\footnote{https://users.ece.cmu.edu/~koopman/ul4600/index.html}
    \item NIST AI RMF\footnote{https://nvlpubs.nist.gov/nistpubs/ai/nist.ai.100-1.pdf}
    \item FDA: Guidance \footnote{https://www.fda.gov/medical-devices/software-medical-device-samd/artificial-intelligence-and-machine-learning-software-medical-device}
    \item ISO/IEC TR 5469:2024 - Artificial intelligence — Functional safety and AI systems \footnote{https://www.iso.org/standard/81283.html}
    \item Aerospace: EUROCAE ER-027 / SAE AIR6987: Artificial Intelligence in Aeronautical Systems: Taxonomy \footnote{https://www.eurocae.net/news/posts/2024/december/eurocae-and-sae-international-publish-landmark-ai-taxonomy-document-for-the-global-aeronautical-industry/}
    \item Aerospace FAA AI Safety Assurance roadmap \footnote{https://www.faa.gov/media/82891}
\end{enumerate}

\subsection{New challenges and opportunities}
In the context of defining what safety is, the participants indentified the following research directions:

\paragraph{Specification languages:} What are appropriate languages that enable capturing context-dependent and rich safety specifications? Formal methods the already enable that in some contexts, but not all; for example, (i) how can one define a language  that includes both absolute and empirical/statistical definitions, (ii) how can one capture abstract properties such as perceived safety, and (iii) how can one define safety properties over semantic and non semantic inputs and outputs.

\paragraph{Benchmarks and resources:} Since safety is context dependent, creating a community resource in the form of competitions or benchmarks that contain a variety of safety problems and metrics would enable researchers from different communities to more easily share ideas and approaches.

%% file: design.tex
As discussed in the previous section, safety is an overloaded and multifaceted term in AI, with context-dependent interpretations and implications. What safety means and how it is best achieved varies greatly depending on the domain: robustness to adversarial inputs in vision tasks, safe actuation in robotics, reliable inference in healthcare, or trustworthy interaction with humans. Consequently, the design principles for safe AI systems must adapt to these differing contexts, reflecting the specific threats, operational constraints, and expectations unique to each domain. In this section, we discuss current approaches to embedding safety as a design goal alongside accuracy, examine whether safety should be treated as a constraint, loss term, or architectural choice, and highlight opportunities for future research.

\subsection{Design for Safety of Machine Learning Systems}

In machine learning systems, safety is often equated with robustness to input perturbations, ranging from adversarial attacks in vision systems, to prompt jailbreaks in large language models (LLMs), to impersonation threats in speech recognition. The prevailing design trend frames safety through robust training objectives: adversarial training, distributionally robust optimization, and learning subject to robustness or safety constrains. Complementary to these are analysis tools like Lipschitz-constrained learning, certification methods (e.g., randomized smoothing, interval bound propagation), and formal verification.

However, there is a growing recognition that many of these robustness strategies operate in narrow and artificial threat models. For example, adversarial attacks to computer vision systems are often restricted to $\ell_p$-norm bounded perturbations, while practical perturbations are due to rain, fog, or man-made modifications to the scene. Likewise, jailbreaking attacks to LLMs remain limited to the hand-crafted design of jailbreaking prompts. Another pressing need is open-set robustness—the ability to generalize safety guarantees to threats not seen during training. Additionally, there is increasing interest in bridging robustness and explainability, as safety failures often stem from unintelligible model behavior.

\subsection{Design for Safety of Control Systems}

In control systems—such as those used in robotics, aerospace, and autonomous vehicles—safety has historically been framed in terms of forward invariance of safe sets, with system behavior guaranteed to remain within these bounds under nominal conditions. Regulatory agencies like the FAA or TSA often serve as certifying bodies. Modern learning-based controllers challenge this paradigm, introducing components (e.g., vision systems or neural network policies) that are difficult to formally verify.

Key design challenges include moving beyond known safe sets to anticipating and responding to unknown, out-of-distribution events. Control systems must be adaptive, making safety-critical decisions in response to novel threats without explicit pre-specification. There is also a need for new certification paradigms that can evaluate systems where perception and control are deeply intertwined—for example, visual input guiding aircraft autopilots. Importantly, safety must also integrate human-in-the-loop expertise, leveraging the judgment of pilots or operators who may override the system under exceptional conditions.

\subsection{Design for Safety of Healthcare Systems}

In the healthcare domain, safety and effectiveness are traditionally assessed through clinical trials and statistical validation, which are certified by regulatory agencies such as the FDA. However, existing methods were designed for structured, low-dimensional data, and are often inadequate for AI systems trained on high-dimensional, constantly evolving datasets. 

A significant gap lies in developing statistical tools that support model retraining as new patient data arrives, while still preserving rigorous evaluation standards. Designing safe AI for healthcare also demands explainability, as clinicians must trust and understand model decisions. Privacy is another essential axis, with risks such as model inversion, membership inference, and data leakage. Finally, clinical AI systems must account for individual variation: each new patient might be out of distribution when compared to the training data, requiring models to reason about uncertainty and applicability on a per-case basis.

\subsection{Design for Safety of Human-AI Interaction}

Human-AI interaction introduces unique safety challenges. Designers must assume that users will behave unpredictably—making errors, gaming systems, or employing them in unintended ways. Accordingly, safety must be proactive and resilient, ensuring that regardless of user behavior, the AI system does not cause harm. This is particularly salient in robotics, where physical safety is at stake, but also applies to LLMs, recommender systems, and medical devices.

Current approaches emphasize safe defaults and fail-safe modes, yet open questions remain. How do we design systems that degrade gracefully under failure? How do we define and measure safety in interactions involving trust, autonomy, and co-adaptation? Moreover, safety is not always a scalar; it can be a multi-criteria objective with conflicting or incomparable tradeoffs. Addressing this requires new frameworks that can reason over partial orders, prioritize safety objectives dynamically, and handle incompatibility between goals.

\subsection{Opportunities}

Designing for safe AI is a complex, multi-domain challenge. Across all domains, there is an urgent need to treat safety as more than an afterthought or static constraint—it must be a first-class design objective, embedded into the core of learning, architecture, evaluation, and deployment. Promising research directions include unifying robustness with explainability, enabling adaptive safety under distribution shift, and developing certification methods that keep pace with learning-based systems. Above all, safety must be contextualized—not just in terms of threats, but in terms of users, environments, and evolving system behavior.

%% file: analysis.tex
\subsection{Current  Techniques and Their Limitations}

We begin by highlighting current techniques and challenges for the safety analysis of AI systems.
Testing and simulation are widely used, but they provide no formal guarantees and can face issues such as manual effort in labeling test data, lack of adequate coverage metrics, or lack of relevance to real-world deployment settings.
Formal verification, on the other hand, provides guarantees, and there are many tools available; however, it suffers from serious scalability challenges and it often cannot deal with semantic properties, which typically do not have an analytical form.
Probabilistic and statistical verification are more natural, due to the uncertainty that arises from the models, but can only provide probabilistic guarantees.
Run-time monitoring can be used to observe and check system behavior during operation, but is limited by the lack of ground truth at run-time. See \cite{dalrymple2024guaranteedsafeaiframework,DBLP:conf/birthday/MitraPPSMLWGY25,10.1145/3503914} for recent surveys on safety analysis for AI systems.

Moral concerns are less studied, but should be central to a safety analysis. 
Defining what constitutes {\em safety} is particularly challenging in the context of autonomous systems, healthcare, or military applications. The concept of safety might mean different things in different settings, and it is important to incorporate ethical considerations into the safety analysis of AI. Incorporating physics into AI models may be a way to prevent models from producing erratic outputs, or ``white noise'', by grounding them in physical laws or principles.

\subsection{Formal guarantees} 
In terms of formal guarantees, several types are possible, including worst-case guarantees, probabilistic guarantees, and confidence intervals; counterexamples are valuable artifacts of formal verification and falsification techniques, while run-time techniques provide complementary assurance.

\paragraph{Worst-case Guarantees:} Worst-case guarantees can be obtained with formal verification tools and provide strong assurance that the analyzed system behaves safely, even in the most extreme, worst-case scenarios. 
Example properties that can be checked formally  include  {\em local robustness}, i.e., invariance of a neural network output with respect to small, norm-bounded perturbations, or {\em input-output properties}, as in the popular ACAS Xu benchmark \cite{10.1007/s10703-021-00363-7}
--
a family of networks designed to prevent mid-air collisions between unmanned aircraft systems (UAS) and other aircraft. 

The annual International Neural Networks Verification Competition (VNN-COMP) \footnote{\url{https://sites.google.com/view/vnn2025}} provides a forum for researchers to evaluate their neural network verification tools on an increasing set of standardized benchmarks, to enable progress within the domain and to understand current limitations. 
However, these types of guarantees are generally difficult to obtain due to scalability challenges and the need to make assumptions about the environment.

\paragraph{Probabilistic Guarantees:} Probabilistic guarantees are more feasible for complex systems, and they more naturally address uncertainty in the environment. Probabilistic properties state that desired properties of a system hold with certain probability  and can be checked using probabilistic verification or statistical, sampling-based techniques.  However, they do not provide absolute guarantees about safety, so they may miss {\em rare}, low-probability events. Confidence intervals can give a range of possible outcomes with a defined level of confidence, strengthening the probabilistic results. 

These techniques rely on assumptions about input distributions, which may not hold in practice; furthermore, the methods are brittle in handling distribution shifts.
It is also difficult to integrate
probabilistic or statistical guarantees when reasoning about larger systems, particularly when those systems involve both
AI and traditional software.
ML systems can make mistakes, even with low probabilities, which complicates the integration of probabilistic guarantees into
systems that rely on traditional verification methods.

\paragraph{Runtime Assurance:}
Complementary runtime verification techniques offer formal assurance through monitoring executions of a deployed learning-enabled system against desired safety properties, often specified in temporal logics. Shielding techniques can be used to guide learning and further enforce such properties at runtime. Further, safety guardrails can be used to check or enforce conditions on the inputs or outputs  of a machine learning model, to detect and mitigate different types of risks.

\paragraph{Testing and Falsification Techniques:} While testing and simulation can not provide formal guarantees, they remain crucial for the reliability and safety of critical systems, where failures can have severe consequences. Although various techniques have been developed to create test suites, requirements-based testing for DNNs remains largely unexplored while adequacy metrics are still missing. Various falsification and heuristic-search techniques combine the power of formal methods with the efficiency of testing techniques, with the goal of finding rare scenarios that lead to critical violations in cyber-physical systems.

\subsection{New challenges and opportunities}

Safety analysis for AI systems is challenging due to the scalability of formal methods, the uncertainty in AI models and environments, and the difficulty in defining safety in diverse applications. Several areas of research have been identified by the workshop participants to address these challenges:

\paragraph{Neuro-symbolic programming:} One notable area is neuro-symbolic programming\footnote{\url{https://neus-2025.github.io/}}\footnote{\url{https://www.nsf.gov/events/neurosymbolic-systems-trustworthy-ai}}, which seeks to combine the strengths of neural networks and symbolic reasoning, to enable explainability and verifiability~\cite{ChaudhuriEPSSY21}. 

\paragraph{Trustworthy AI agents:} The concept of trustworthy AI agents was also raised, highlighting the need for agentic AI systems that are reliable, predictable, and aligned with ethical
standards. Trustworthy AI agents can be realized by integrating formal verification in the agentic workflow, ensuring that the agents behavior adheres to predefined specifications, and detecting potential errors or vulnerabilities. 

\paragraph{Multi-modal and large language models:} Multi-modal models that combine different types of data (e.g. vision,
text, and speech) are another promising avenue for improving the specification and control of system properties. For instance, one can use the text modality to define and check natural-language requirements on the image modality. The use of Large Language Models (LLMs) to probe embedding spaces was suggested as a way to test how AI models handle complex, high-dimensional data and uncertainties. The possibility of using LLMs to elicit requirements, generate proof objects or create diverse scenarios for training and testing purposes was also discussed as promising avenues for future research. 

\paragraph{Benchmarks:} The participants also pointed out the need for realistic testbeds and benchmarks\footnote{\url{https://trustllmbenchmark.github.io/TrustLLM-Website/}}. Further, the participants emphasized the need for new dynamic benchmarks and verification methods that can reflect the complexities of the real world, as traditional verification often assumes static, unrealistic environments.  

\paragraph{Human-AI collaboration:} Human-AI collaboration was emphasized as essential, especially in fields such as cybersecurity, where AI could help identify vulnerabilities.
AI systems should be trained in line with both statistical and worst-case guarantnees, but there is also the concern that overapproximating worst-case scenarios could lead to systems that are overly cautious and consequently could be less safe.

\paragraph{Other challenges and opportunities:} Other challenges include the lack of precise model assumptions, which can lead to system failures if the assumptions are not valid in practice. Making these assumptions explicit is crucial to improving the reliability of the system. Adversarial robustness is another issue, as AI systems can often fail when exposed to previously unseen attacks, even if they perform well on known ones. Defining what constitutes ``safety" in AI systems is still a major challenge. Safety definitions need to be clearer, especially in fields such as autonomous driving or medical devices, where the potential for harm is high. These issues are described in depth in the previous sections, but are also very relevant for safety analysis. Finally, the participants believe that close collaboration between the ML and verification communities offers promising avenues forward.





%% file: llm.tex
An emerging body of work has focused on attacks on and defenses for generative AI systems specifically, particularly large language models (LLMs), vision-language models (VLMs), and AI agents, whether embodied or virtual. We believe that these merit their own discussion given the outsize role they occupy in discussions of safety.

\subsection{Overview of Concepts}

We highlight important concepts for defining the scope of attacks and defenses on current AI systems. First, {\it what is the goal of the attack?} Is it to compromise a model's performance, a model's fairness, a models' explainability, user data or privacy, or something else? Second, {\it what domain of model is being attacked?} Is it a large language model (LLM), a vision-language model (VLM), a foundation model for robotics, a code model, a tool-calling model, or a multi-agent system? Finally, {\it what is assumed about the system being attacked?} Do we assume white-box or black-box access?

Within these questions, several aspects of attacks and defenses emerge specifically. These fall under the category of ``Safety with respect to adversaries'' defined in Section~\ref{sec:types_of_safety}.

\paragraph{Attacks} There are many categories of attacks. A first is prompt injection or jailbreaking attacks, which attack a model's performance at inference time \cite{Yi2024JailbreakAA}. These attacks involve an adversary circumventing safeguards in place to get LLMs to perform potentially unsafe actions from the perspective of the system developer. Second, data poisoning attacks involve changing training data to induce unsafe behavior in a model, such as injecting a trigger to enable some unsafe behavior by changing a small part of the LLM's training data \cite{Bowen2024DataPI,He2024DataPF,Bowen2025ScalingTF}. Finally, model extraction or data extraction attacks involve attacking privacy of either developer model parameters to understand how the model was built \cite{Carlini2024StealingPO} or extracting sensitive user data that the model may have been trained on \cite{Patil2023CanSI}. Some of these are well-studied, but we believe they are understudied in the areas of tool use and robotics foundation models, where new applications of foundation models lead to new possible threats.

\paragraph{Defenses:} There are many categories of defenses. Defense starts with red-teaming, which enables identifying failures in the failure by probing possible failure modes \cite{Mazeika2024HarmBenchAS}. One defense is adversarial training, which defends the model against possible failure modes by specifically inoculating for those failures, or just generally improving robustness to perturbations of prompts \cite{Xhonneux2024EfficientAT,Sheshadri2024LatentAT}. An attractive approach is to bound neural network behavior formally \cite{Jia2019CertifiedRT}; however, it is difficult to achieve strong bounds for LLMs in practice. Finally, we can ``work around'' failure modes by implementing neurosymbolic approaches or building models into pipelines, which isolate any potential harm from the model in the overall system. A related idea is to educate users of a model, so when they see potentially erroneous or dangerous outputs, the harm is minimized; this approach isolates society itself from the model's ill effects. Several of these defenses can be improved through model interpretability, or understanding how a blackbox neural network model functions internally.

Within this space of topics, we highlight four topics that we think are particularly worthy of future study. These represent either emerging areas of attacks and defenses on cutting-edge applications or understudied aspects of the attack-defense landscape. They are not meant to preclude the value of other research along the directions mentioned above.

\subsection{Agentic, tool-use, robotic system vulnerabilities}

Models that interact with external systems or with the real world pose additional threats beyond systems that just generate text or images. A model could be attacked either through prompt injection to trigger undesirable behavior or through manipulation of its training data to include a backdoor. Defending against such malicious usages requires guaranteeing the API calls are safe. How do we define safety, how do we guarantee it, and is it conceivably feasible?

Although prompt injection is well-studied, we believe the idea of {\it data poisoning} for these systems is underexplored. For instance, if training data is manipulated such that (a) a robot can receive a special signal that tells it to do something bad; or (b) injection from Internet data into a VLM causes actual downstream harm in some system.

In this space, a major gap is {\it benchmarking}. Some studies exist \cite{chen2024agentpoisonredteamingllmagents} but we believe more should be done. What should a benchmark look like when thinking about either robots or agents and ensuring safety with respect to LLMs composing these APIs? This could require defining a canonical architecture for an LLM/VLM/foundation model and a platform or API it interacts with. What are the representative examples or use cases? How do we build up simulators to explore this? Are they reusable across groups without access to particular robotic platforms?

\subsection{Applications of interpretability}

Interpretability is an intriguing path towards defending AI systems. At its core, the idea is that understanding these systems may enable us to mount more effective defenses. We envision several ingredients here.

First, continued work on {\it analyzing LLM internals} can be useful. For instance, by identifying precise ``circuits'' that get activated during the forward pass, we can understand what inputs might activate harmful behaviors. An attacker could use this knowledge to do some kind of prompt injection in a more targeted way: we know what capability in the model we want to ``derail'', and understanding the model lets us derail it more effectively. But conversely, the same understanding may enable defenses as well.

Second, this work can be broadened to consider {\it human-in-the-loop} systems. When we use interpretability as a way of analyzing systems for their potential to be attacked, very expert humans are still required to do this. Is there a way for domain experts to be able to use interpretability tools and understand the shortcomings or attack surfaces of their models/applications?

Across these ideas, challenges include: (1) {\it generalizability of findings} across models, datasets, tasks, and domains. (2) {\it the need for more mature interpretability approaches}, as current studies are sometimes limited to toy examples and applications.

\subsection{Multi-agent foundation model systems}

One emerging area is multi-agent foundation model systems. One example of this so far is {\it cooperative} groups of agents. For instance, ensembles of models or models that use separate verification and correction processes can both be construed as multiple models cooperating. What attacks and defenses are possible in the context of such systems? How does the robustness of these systems compare to monolithic systems? Although these workflows are currently largely static, we expect this to change as the constituent agents become more powerful and more independent. For instance, groups of agents could collaborate to figure out their own strengths and weaknesses and defend against attacks. They might also need to mitigate ``overpersuasion'', or being convinced by an erroneous agent. Robustness against attacks emerges from knowing which agent can handle which task best. Such multi-agent systems may also be more robust against data poisoning attacks

We can also consider {\it competitive} or {\it independent} agentic systems. For instance, in a world where users each have LLM agents that are negotiating with each other to find time for meetings, the incentives of these systems are not all aligned. A rogue actor in this environment could inject unwanted behavior into the system by behaving in a poor way. Conversely, a group of agents could band together to mitigate such poor behavior, particularly in non-zero-sum settings.

We see intriguing connections between this kind of work and information ecosystems among humans: do they propagate or attenuate misinformation? 

\subsection{Long-term autonomous LLM systems}

Finally, we consider the rise of ``long chain-of-thought'' systems like DeepSeek-R1 as an opportunity for further research. These systems are bridging from single-turn interaction (ChatGPT) to long-term interaction with the world, including revision of beliefs and lifelong learning within a single inference trace. We believe these represent a new frontier of systems. When should models make parameter updates vs.~summarize what they’ve figured out so far as a discrete memory? How could attacks on these models develop as research matures? Is their reasoning more robust or less robust to attacks than other reasoning models?